\documentclass
[aps,prl,noshowpacs,groupedaddress,fleqn,fleqnarray,amsmath,amssymb]{revtex4}

\usepackage{graphicx,hyperref}

\usepackage{amssymb}
\usepackage{epstopdf}
\usepackage{url}

\usepackage{bm}
\usepackage{graphics}
 
\def\beq{\begin{equation}}
\def\eeq{\end{equation}}
\def\beqar{\begin{eqnarray}}
\def\eeqar{\end{eqnarray}}

\def\0{\mbox{\boldmath$\displaystyle\mathbb{O}$}}

\def\1{\mbox{\boldmath$\displaystyle\mathbb{I}$}}

\begin{document}
\noindent
{\Large \sc Neutrino oscillations: Inevitability of non-standard interactions or a sterile neutrino}

\vspace{21pt}\noindent
{\sc D. V. Ahluwalia and D. Schritt}
\vspace{11pt}

\noindent
Department of Physics and Astronomy,  Rutherford Building\\
University of Canterbury, 
 Private Bag 4800, 
Christchurch 8020, New
Zealand

\vspace{11pt}

\noindent 
E-mail: dharamvir.ahluwalia@canterbury.ac.nz, dsc35@student.canterbury.ac.nz
\vspace{21pt}


\vspace{21pt}

\maketitle

\noindent\textbf{Insistence on the conservation of energy led Pauli to conjecture the existence of neutrinos. The data that has since accumulated adds a new twist to the story.  We now know that not only do neutrinos appear in three different flavours, but each of these is a linear superposition of three different mass eigenstates.  
Neutrinos are thus truly quantum mechanical particles with no classical counterpart. They neither carry a definite mass, nor a definite energy. This makes them oscillate from one flavour to another. In general, each  flavour has a different
energy expectation value. Neutrino oscillations thus raise a fundamental question: how is energy conserved when a neutrino of one flavour oscillates to another flavour? The energy imparted in the flavour measurement process is not sufficient to explain the resulting paradox as,  by choosing  an appropriate kinematical setting,   this `back reaction' can be made to average to zero. To resolve this paradoxical situation, we here propose a conjecture that the neutrino mixing matrix must be such that in the oscillation process the energy expectation values  enjoy flavour independence. This conjecture turns out to be very  powerful. Using  the existing data, it suggests that either neutrinos have interactions that go beyond those expected from the standard model, or there must exist sterile neutrinos.
}
\vspace{11pt}

Thus we present unitarity and energy conservation constraints on neutrino oscillation parameters.  In our analysis  ${\theta_{13}}$ depends on 
the ratio of the mass-squared differences $\zeta$ and on  ${\theta_{12}}$ alone; while ${\delta_{CP}}$ is completely determined by  $\zeta$,  ${\theta_{12}}$, and  ${\theta_{23}}$.  The standard neutrino oscillation phenomenology (SNOP) suggests ${\zeta \approx 30}$. The resulting prediction for ${\theta_{13}}$ conflicts with SNOP expectation for all values of ${\theta_{12}}$. This leads to the suggestion that  some non-standard neutrino interactions may be intervening to invalidate the SNOP inferred ${\zeta}$ and ${\theta_{13}}$. An alternate inference is to go beyond the three flavours of SNOP and incorporate sterile neutrinos in the analysis.

\vspace{11pt}

In this communication we shall confine to the simplest, and the standard, neutrino oscillation phenomenology. This approach allows us to keep all our calculations analytical and it helps us to retain an element of conceptual minimalism. The justification for such an approach resides in the linearity of the Hilbert space structure  of quantum mechanics. 
We adopt the standard parameterisation of mixing matrix $U$. Thus, it is  characterised by three angles $\theta_{12}$, $\theta_{13}$, and $\theta_{23}$ and a CP violating phase $\delta_{CP}$~\cite{Amsler:2008zzb}. We define the neutrino flavours as usual
$
\vert\nu_\ell\rangle = \sum_i U_{\ell i}  \vert\nu_i\rangle
$
 with  the flavour index $\ell = e,\mu,\tau$, and mass eigenstate index $i = 1,2,3$.  In what follows we shall need to
evaluate expectation values of energies for each of the three flavours. We shall assume that a  flavour measurement  
does not change the momentum of the underlying mass eigenstates.
 \vspace{11pt}

To be explicit, consider an electron neutrino; or, more precisely an ensemble of similarly produced electron neutrinos. If a large number of energy measurements
are performed, then on the average the result is the expectation value  $\langle E_e \rangle = \langle \nu_e\vert H \vert \nu_e\rangle= \sum_i U^\ast_{e i} U_{e i} E_i $, where $E_i = p + m_i^2/(2 p)$. But one may wish to forgo energy measurements, and allow the 
electron neutrino to evolve. Then at a later time a flavour measurement is made, and, say, the measurement yields a muon neutrino. 
In large number of similar measurements, and immediately after a muon neutrino is detected, the average result in an
  individual measurement is $\langle E_\mu \rangle =\sum_i U^\ast_{\mu i} U_{\mu i} E_i $.
 These
expectation values are accompanied by fundamental uncertainties 
  $\Delta E_\ell = \pm \sqrt{ \langle\nu_\ell \vert H^2 \vert
   \nu_\ell\rangle - \langle \nu_\ell\vert H \vert \nu_\ell\rangle^2} $ with
   $H\vert \nu_\ell\rangle =  \sum_i E_i U_{\ell i} \vert\nu_i\rangle$.
\vspace{11pt}

Now consider an idealised beam of neutrinos, initially containing electron neutrinos only.
The state of a single electron neutrino $\vert\nu_e\rangle=\sum_i U_{e i} \vert \nu_i\rangle$ evolves with time, and at any given time it has a definite probability of being found \textemdash~ on measurement \textemdash~ as one of the three neutrinos, $\vert\nu_\ell\rangle$. Unless a flavour measurement is made the energy expectation value remains constant in time. However, if a flavour  measurement is made the
initial energy content $ \mathcal{E}(t=0) :=  N   \langle E_e \rangle$, at any later time $t$  gets distributed into three components
\begin{equation}
 \mathcal{E}(t>0)= \sum_\ell 
 \underbrace{N P(\nu_e\to\nu_\ell) \langle E_\ell\rangle}_{\mbox{contribution from flavour}~ \ell}
 \end{equation}
where $N$ represents the number of electron neutrinos at the initial time $t=0$ and $P(\nu_e\to\nu_\ell)$ is the oscillation probability
to flavour $\ell$ (at time $t$).
If we restrict the mixing matrix $U$  such   that all three  $\langle E_\ell\rangle$ are equal, the conservation of energy  requirement translates to a simple condition of unitarity: 
 $\sum_\ell P(\nu_e\to\nu_\ell) 
   =1$.   
Our central conjecture is that given a $p$,  the three $\langle E_\ell\rangle$ must be  flavour independent.  
 \vspace{11pt}

 The usual arguments~\cite[Appendix C]{Ahluwalia:1996fy} that are extended in this context fail~\cite{Haines:To1998} because the time associated with the measurement of a flavour $\Delta t_\ell \ll 
 c/\lambda_{ij}$, and not  of the order of ${c}/
 {\lambda_{ij}}$. Here $\lambda_{ij}$ is the oscillation length associated with the mass squared difference $\Delta m^2_{ij}$, and $c$ is the speed of light~\footnote{We thank Todd Haines (Los Alamos) for bringing this to our attention.}.

 \vspace{11pt}  
For a general mixing matrix our conjecture on unitarity and energy conservation cannot be satisfied. But, as we will see below, there exists a class of mixing matrices for which the conjecture does hold. Here we argue that when the resulting mixing matrices are seen in the light of existing data one is forced to infer the existence of a sterile neutrino, or/and that  neutrinos engage in new interactions.
\vspace{11pt}

The  conjecture is implemented by seeking a solution to the following two equations:   $\langle E_e\rangle -   \langle E_\mu\rangle=0$ and  $\langle E_\mu\rangle -   \langle E_\tau\rangle=0$, which together imply $\langle E_e\rangle -   \langle E_\tau\rangle=0$.  Given the mixing angles $\theta_{12}$ and 
 $\theta_{23}$, and a parameter $\zeta$ defined by the the mass-squared differences
\beq
\zeta:= - \left(\Delta m^2_{32}/ \Delta m^2_{21}\right)
\eeq
the CP-violating phase $\delta_{CP}$, and  the mixing angle $\theta_{13}$, are constrained to be:

\begin{eqnarray}
&& \theta_ {13} = \eta_1
 \arccos\left[\eta_2 \sqrt{
 \frac{2}{3} \left(
 \frac{2\zeta -1}
 {2 \zeta - \cos\left(2 \theta_{12} \right) -1}\right)
}  \right]\label{eq:t13} \\
&&  \delta_{CP} = \eta_3 \arccos \left[\eta_4
\Big( 2\zeta + 3 \cos\left(2 \theta_{12}\right) -1\Big)\frac{\cot\left( 2 \theta_{23} \right)}{\sin\left(2\theta_{12}\right) }
\sqrt{ \frac{1}{3} \left(\frac{2\zeta - \cos\left(2 \theta_{12}\right) -1}{2\zeta - 3 \cos\left(2 \theta_{12}\right) -1}\right)
  } \right]  \label{eq:dcp}
\end{eqnarray}
where the $\eta$ are either $+1$ or $-1$.
The set $\{\eta_1,\eta_2,\eta_3,\eta_4\}$ 
contains sixteen elements. The set containing the signatures $\{ -,-,-,-\},
\{-,-,+,-\}, \{+,-,-,+ \}, \{+,-,+,+ \}, \{-,+,-,- \}, \{-,+,+,- \}, \{+,+,-,+ \},
 \{+,+,+,+ \}$ allows for all values of $\zeta$, $\theta_{12}$, and
 $\theta_{23}$ provided the indicated `$\arccos$' exists. The remaining set of eight signatures requires $\theta_{23}$ to be $\left(n\pm \frac{1}{4}\right)\pi$ with $n$ an integer; or $\theta_{12}$ to be 
$\frac{1}{2}\arccos\left(\frac{1-2\zeta}{3}\right)$. Solutions (\ref{eq:t13}) and (\ref{eq:dcp}) are defined modulo an additive factor of $2 n \pi$.

\begin{figure}
\includegraphics[width=400pt]{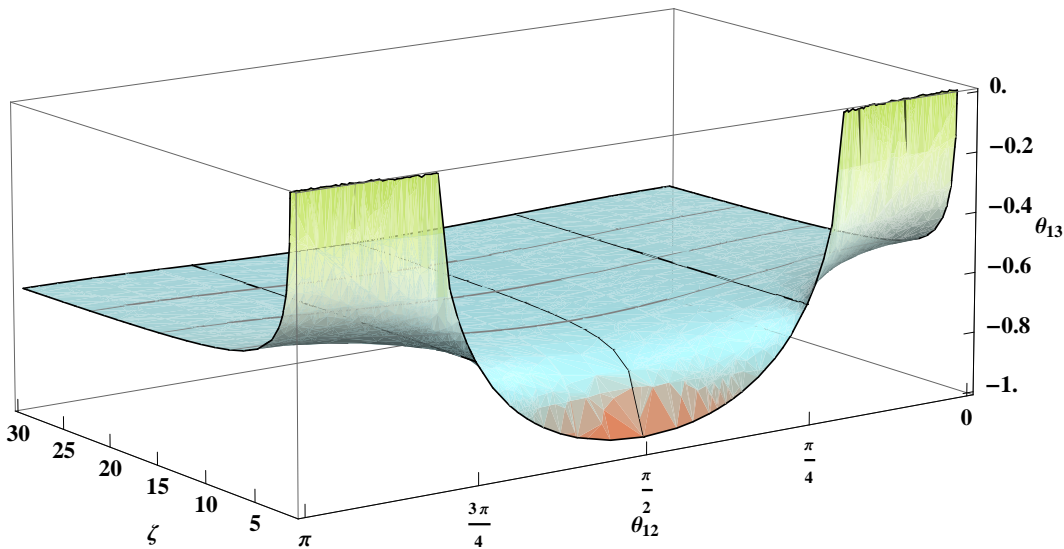}
\caption{The variation of  $\theta_{13}$ as a function of $\zeta$ and $\theta_{12}$. Signature of the plotted solution is $\{-,+,+,-\}$.}
\end{figure}

\vspace{11pt}
Within the framework of the SNOP, 
a careful analysis of the existing data, assuming CP conservation ($\delta_{CP}=0$) and the standard MSW effect,  yields~\cite{Escamilla:2008vq} 
\begin{equation}\theta^{snop}_{13} = -0.07 ^{+0.18}_{-0.11}\label{eq:data13}
\end{equation}
 at the $90\%$ confidence level. The sign of $\zeta$ is not known. We shall take $\zeta>0$, and infer $\zeta^{snop} \approx  30$ from the analysis given in Ref.~\cite{GonzalezGarcia:2007ib}.  
 \vspace{11pt}

 Taking a hint from the analysis given in~\cite{Escamilla:2008vq} we here  examine the arrived constraints with the signature 
 $\{-,+,+,-\}$.
In Figure 1 we have plotted $\theta_{13}$  given by Eq.~(\ref{eq:t13}) as a function of $\zeta$ and $\theta_{12}$.
 After taking note of the fact that $\theta_{13}$ in Eq.~(\ref{eq:t13}) does not depend on $\delta_{CP}$ and $\theta_{23}$,
 we immediately see that if  $\theta_{13}$ is given the value shown in Eq.~(\ref{eq:data13}) then $\zeta \approx 1 $, and not $30$.  
In SNOP, with the exception of short base line experiments, the MSW effect plays a crucial role, and for that reason the inferred value of $\zeta$ sensitively depends on the assumption that neutrinos engage in standard model interactions. The arrived conflict strongly indicates that this assumption may not be true; or, that to relax the constraint on $\theta_{13}$ given by Eq.~(\ref{eq:t13}) one must go beyond the three neutrino flavours and consider the possibility of a sterile neutrino.
\vspace{11pt}

An analysis with $\zeta<1$, and for other signatures of the $\eta$ metric, has been done. The essential conclusion remains unchanged.
\vspace{11pt}

This leads us to suggest a series of short base line neutrino oscillation experiments where matter effects may be safely ignored.
Freed from matter effects, these experiments can then test if the constraints that we arrive at are consistent with observations or if a sterile neutrino is needed.
At the same time, we advise that the existing  neutrino oscillation data be re-analysed within a framework that allows neutrinos to carry non-standard model interactions (NSI)~\cite{Grossman:1995wx,Gavela:2008ra,Guzzo:2004ue,Escrihuela:2009up,Ohlsson:2009vk,Biggio:2009nt,Minakata:2009gh,Mitsuka:2009zz,Gago:2009ij,Meloni:2009ia,Kopp:2008ds,Davidson:2003ha,Huber:2001zw,Nunokawa:2000ff,Bergmann:2000gp,Fornengo:2001pm,Koike:2000jf}
\emph{and} that takes into account the constraints given in  equations (\ref{eq:t13}) and 
(\ref{eq:dcp}).
\vspace{11pt}

In the context of recent works on NSI two generic remarks may be kept in mind: (a) in phenomenological approaches the parameter $\zeta$, or its equivalent, is usually taken to be small; and (b) the Elko reduction of mass dimensionality of spinor fields is completely ignored~\cite{Ahluwalia:2004ab,Ahluwalia:1996fy}. To examine our proposal experimentally both of these assumptions need to be abandoned. In addition, one may wish to include neutrino-dark matter interactions.
This may be particularly important in the context of the solar neutrino data.  
\vspace{11pt}

\begin{figure}
\includegraphics[width=400pt]{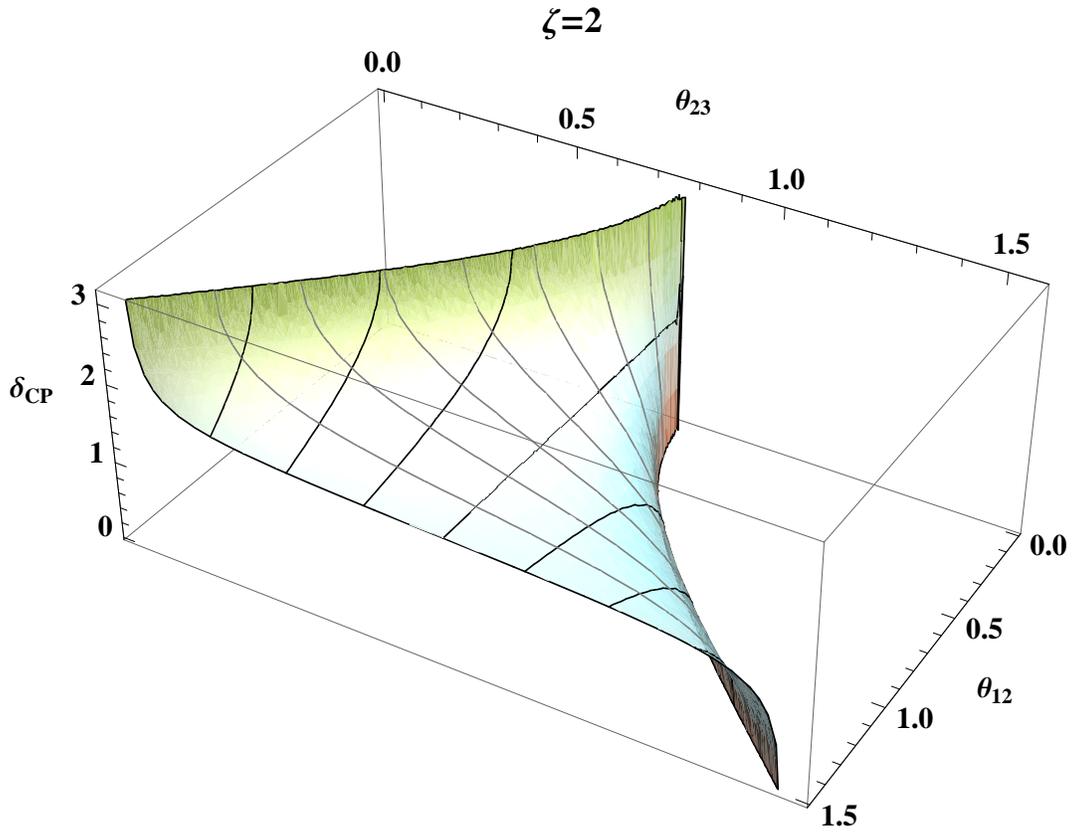}
\caption{\label{Fig:2} The variation of  $\delta_{CP}$ as a function of $\theta_{12}$ and $\theta_{23}$ for $\zeta=2$ . Signature of the plotted solution is $\{-,+,+,-\}$.}
\end{figure}

\begin{figure}
\includegraphics[width=400pt]{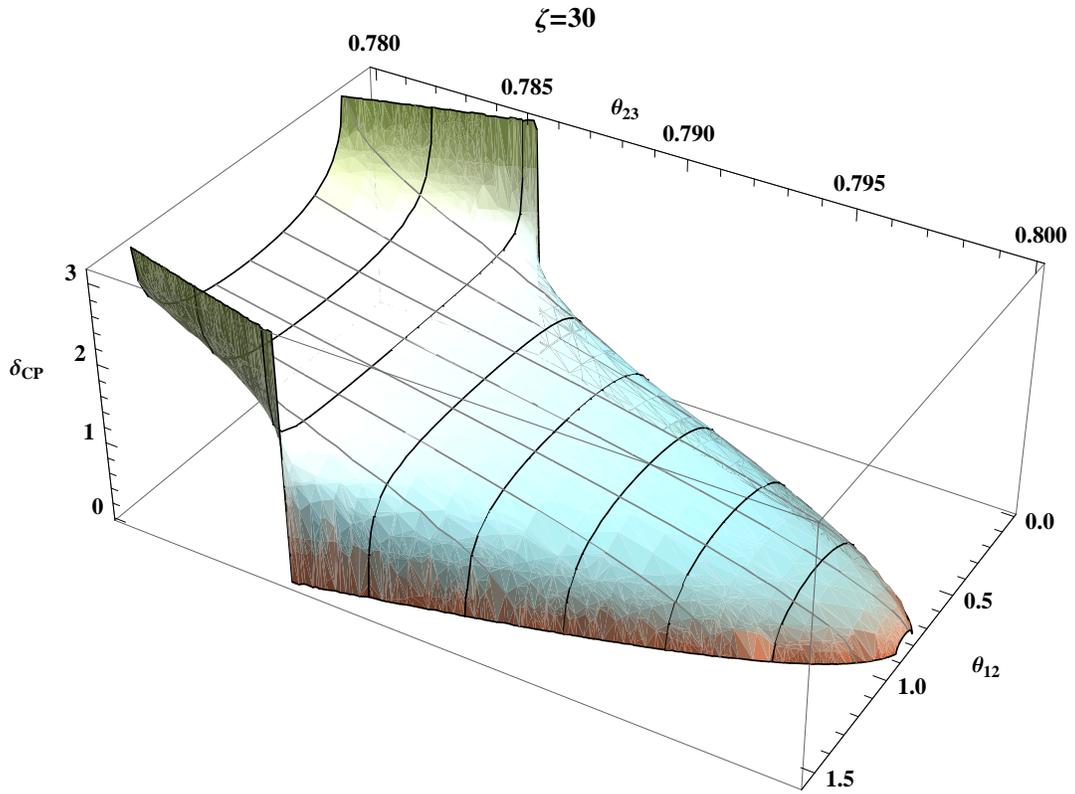}
\caption{\label{Fig:3} The variation of  $\delta_{CP}$ as a function of $\theta_{12}$ and $\theta_{23}$ for $\zeta=30$. Signature of the plotted solution is $\{-,+,+,-\}$.}
\end{figure}

Combined, these two efforts shall place our knowledge about neutrino oscillations on a firmer ground while at the same time allow us to examine the predictions that we make here. Assuming three neutrino flavours, our constraints predict that the CP violating phase  $\delta_{CP}$  can be determined from an accurate measurement of $\zeta, \;\theta_{12}$ and $\theta_{23}$. Similarly, a determination of $\zeta$ and $\theta_{12}$ alone suffices to predict the value of $\theta_{13}$. This is graphically reflected in Figures \ref{Fig:2} and \ref{Fig:3}. The complete information is contained in equations (\ref{eq:t13}) and 
(\ref{eq:dcp}). The argument that brought neutrinos on the theoretical scene now tells us how neutrinos oscillate.
\vspace{21pt}

\noindent{\textbf{Acknowledgments.}}
We thank Cheng-Yang Lee for his involvement in the early stages of this work.
\vspace{11pt}

\noindent{\textbf{Author contributions.}} All authors contributed equally to this paper. D.V.A. advanced the conjecture, and with D.S. deciphered its implications. The first working draft was written by D.V.A.

\newpage

\vspace{21pt}
\url{http://www2.phys.canterbury.ac.nz/editorial/}

\end{document}